\documentclass[11pt]{article}
\newfont{\larom}{cmbx9 scaled\magstep3}
\newfont{\bsan}{cmssbx10}
\newcommand{\be}{\begin{equation}}
\newcommand{\ee}{\end{equation}}
\newcommand{\lb}[1]{\label{#1}}
\begin{document}
\begin{center}
  {\larom Scale Invariance in a Perturbed Einstein-de Sitter
  Cosmology}\\
  \vspace{10mm}
  {\Large Elcio Abdalla$^\ast$, Roya Mohayaee$^\dag$, and Marcelo B.\
          Ribeiro$^\ddag$ }\\
  \vspace{10mm}
  \begin{flushleft}
  $~^\ast$ Instituto de F\'{\i}sica, Universidade de S\~{a}o Paulo - USP,
           CxP 66318, CEP 05315-970, \\ \hspace{3mm} S\~{a}o Paulo,
	   Brazil; E-mail: eabdalla@fma.if.usp.br\\
  $\dag$   \hspace{0.01mm} Dipartimento di Fisica, Universit\'a degli
           studi di Roma ``La Sapienza'' 5, Piazzale \\ \hspace{3mm} Aldo
	   Moro - 100185, Roma, Italy; E-mail: roya@titanus.roma1.infn.it\\
  $\ddag$  \hspace{0.005mm} Instituto de F\'{\i}sica, Universidade do
           Brasil - UFRJ, CxP 68532, CEP 21945-970, \\ \hspace{3mm} Rio
	   de Janeiro, Brazil; E-mail: mbr@if.ufrj.br\\ 
  \end{flushleft}
  \vspace{10mm}
  {\bf ABSTRACT}
\end{center}
\begin{quotation}
  \small 
\noindent This paper seeks to check the validity of {\it the apparent
fractal conjecture} (Ribeiro 2001ab), which states that the observed
power-law behaviour for the average density of large-scale distribution
of galaxies arises when some observational quantities, selected by their
relevance in average density profile determination, are calculated along
the past light cone. Since general relativity states that astronomical
observations are carried out in this spacetime hypersurface, observables
necessary for direct comparison with astronomical data must be
calculated along it. Implementing this condition in the proposed set
of observational relations profoundly changes the behaviour of many
observables in the standard cosmological models. In particular, the
average density becomes inhomogeneous, even in the {\it spatially}
homogeneous spacetime of standard cosmology, change which was already
analysed by Ribeiro (1992b, 1993, 1994, 1995) for a non-perturbed model.
Here we derive observational relations in a perturbed Einstein-de Sitter
cosmology by means of the perturbation scheme proposed by Abdalla and
Mohayaee (1999), where the scale factor is expanded in power series to
yield perturbative terms. The differential equations derived in this
perturbative context, and other observables necessary in our analysis,
are solved numerically. The results show that our perturbed Einstein-de
Sitter cosmology can be approximately described by a decaying power-law
like average density profile, meaning that the dust distribution of this
cosmology has a scaling behaviour compatible with the power-law
profile of the density-distance correlation observed in the galaxy
catalogues. These results show that, in the context of this work, the
apparent fractal conjecture is correct.
\end{quotation}

\section{Introduction}

It has been known since 1970 that the large-scale average density mass
distribution of the Universe, constructed with astronomical data on
the galaxy distribution, decreases linearly with increasing distances,
following a power-law pattern in a log-log plot, and with an exponent in
the range from 1.7 to 2 (Kihara and Saki 1970; de Vaucouleurs 1970, and
references therein). Further observational and theoretical analysis of
this behaviour, made in the 1970's by Peebles and collaborators (see
Peebles 1980, and references therein), interpreted this effect as being
a result of small scale density perturbations necessary for the
formation of structures in the Universe. Since then it has been a more
or less conventional wisdom to suppose that such a power-law density
profile arises as a result of these density fluctuations, an effect
being short range in nature, and doomed to disappear at higher distances
where the homogeneity predicted by the Friedmann cosmological models is
supposed to be observed.

This conventional picture was, however, challenged in 1987 by L.\ 
Pietronero, who, taking earlier suggestion advanced by Mandelbrot
(see Mandelbrot 1983, and references therein), and which was also
discussed by Peebles (1980), proposed that the luminous large-scale
matter distribution in the Universe should follow a scale invariant
pattern, arising from the underlying smoothed-out and averaged fractal
galaxy distribution system.

Pietronero's (1987) paper had the effect of starting a sharp
controversy in the literature between the proponents and opposers of
this ``fractal universe'' (see Turok 1997 and references therein).
This controversy is so far mainly focused on the issues of whether
or not observations of large-scale galaxy distribution support or
dismiss an average density power-law profile that decays at
increasing distance, and the depth of this scale invariant system
(see the reviews by Ribeiro and Miguelote 1998, Sylos-Labini et
al.\ 1998, and references therein). Although, a consensus on these
points is yet to be achieved, it is clear by now that settling, or
even clarifying, those controversial points has become an important
issue in cosmology, inasmuch as this ``fractal debate'' has already
reached the main stream of cosmological research (Turok 1997; Coles
1998; Wu, Lahav and Rees 1999; Mart\'{\i}nez 1999).\footnote{ \ Pietronero's
(1987) article is only the most recent form in which the old idea that
matter in the Universe is structured in a hierarchical manner has
resurfaced. For instance, following de Vaucouleurs' (1970) case
for a hierarchical universe, Wertz (1970, 1971) advanced a model
mathematically identical to Pietronero's (1987), where a discussion
about scaling in galaxy clustering can already be found. However,
as scale invariant ideas had not yet appeared, Wertz was unable to reach
many important conclusions found later, and independently, by
Pietronero, in special the significance of the fractal dimension in
galaxy clustering, and the proposal of statistical tools able to
appropriately describe a smoothed-out hierarchical pattern. That may
explain why Wertz's work has remained largely ignored so far. Unaware
of Wertz's work, but inspired by de Vaucouleurs, Mandelbrot (1983)
revived the hierarchical universe model and made a thorough discussion
about the scaling properties of galaxy distribution, fully characterizing
it as a scale invariant structure. A discussion about the similarities
and differences between Wertz's and Pietronero's approaches to the
problem of universal hierarchical clumping of matter can be found in
Ribeiro and Miguelote (1998; see also Ribeiro 1994).} In any case, at
very short redshifts (for interstellar medium, and clusters of galaxies)
we know that there is a scale invariant structure.

Since this is a debate in cosmology, it is only natural that the
feasibility of some kind of scale invariant universe model should also
be investigated in a relativistic framework, and, therefore, relativistic
aspects of cosmological models bearing average density scaling features
can be expected to play an important role in this debate.

Some relativistic cosmologies of this type have already been proposed
in the Lema\^{\i}tre-Tolman-Bondi (LTB) spacetime, which is the most general
spherically symmetric dust solution to Einstein's field equations
(Ribeiro 1992a, 1993, 1994; Matravers 1998; Humphreys, Matravers and
Marteens 1998), but a potentially important investigation is the possibility of
appearance of scale invariant features in Friedmann cosmologies with small
scales perturbations. If average densities with decaying power-law like
profiles could somehow appear in the standard cosmological models, many
points of the current debate about the density behaviour of galaxy
distribution could be clarified, or even resolved, if a relativistic
perspective is taken for these effects.

In a previous paper, one of us (Ribeiro 1992b) studied observational relations
in a non-perturbed Einstein-de Sitter (EdS) cosmology, that is, without any
type of metric or density perturbation, and the general conclusions
were that this cosmology does not show scaling features along the past light
cone, in the sense of not having a power law decrease of the average density
at increasing distances, as predicted in Pietronero's (1987) original model.
However, since all calculations took the backward null
geodesic into consideration it was clearly demonstrated that this
model does not show up as {\it observationally} homogeneous either,
even at small redshifts ($z \approx 0.04$), and this result is
a consequence of the fact that the homogeneity of the standard cosmological
models is {\it spatial}, that is, it is a {\it geometrical} feature
which does not necessarily translate itself into an astronomically
observable quantity. Although a number of authors are aware of
this fact, what came as a surprise had been the low value for the
redshift where this observational inhomogeneity appears. Therefore, it
was clear by then that relativistic effects start to play an important
role in observational cosmology at much lower redshift values than
previously assumed. Nonetheless, at least part of these scaling features
can be analysed in terms of a purely non-relativistic model, as done in
Abdalla et al.\ (1999), though in such a case direct comparison with
observational data is more difficult.

In a sequel paper (Ribeiro 1995), this result was further analysed and the
reasons why this relativistic effect seems to have been overlooked in the
literature was clarified. Due to the non-linearity of the Einstein field
equations, observational relations behave differently at different
redshift depths. Thus, while the linearity of the Hubble law is well
preserved in the EdS model up to $z \approx 1$, a value
implicitly assumed by many other authors as the lower limit up to
where relativistic effects could be safely ignored, the density is
strongly affected by relativistic effects at much lower redshift
values. A power series expansion of these two quantities showed that
while the zeroth order term vanishes in the distance-redshift relation,
it is non-zero for the average density as plotted against redshift.
This zeroth order term is the main reason for the different 
behaviour of these two observational quantities at small redshifts.
Pietronero et al.\ (1997) referred to this effect as the ``Hubble-de
Vaucouleurs paradox'', however, from the discussion above it is clear
that there is no paradox. Indeed what seems to be a paradox are just
very different relativistic effects on the observables at the moderate
redshift range ($0.1 \le z < 1$).

Similar effects of departures from the expected Euclidean results at
small redshifts were also reported by Longair (1995, p.\ 398), and the
starting point for his findings was the same as that of Ribeiro
(1992b, 1995): the use of source number count expression along the null
cone. Nevertheless, the path followed by Longair was quite different
from Ribeiro's. While the former kept his conclusions essentially
qualitative and did not make further investigations of the consequences
of this effect in other observational quantities, like the two-point
correlation function, or provided an explanation for the underlying 
reasons for this effect, the latter attempted to address all these 
issues (see details in Ribeiro 1995).

Despite these interesting and encouraging results, they must still be
considered as preliminary, inasmuch as the analysis advanced
by Ribeiro (1992b, 1995) was carried out in an unperturbed model, and,
therefore, some of its features are unrealistic, especially the behaviour
of the observational quantities at very small $z$, where the average
density tends to a constant value. In addition, one important question
remained. While Ribeiro (1992b) showed that an unperturbed EdS
model does not have scale invariant features, in the sense of not having
a power law decay of the average density at increasing depths, it,
nevertheless, also showed very clearly that {\it there is indeed} a
strong decay of the average density at increasing values of the
luminosity distance or the redshift, an effect termed by Ribeiro
(1995) as ``observational inhomogeneity of the standard
model''.\footnote{ \ Note that this result is achieved only when
an appropriately chosen set of observational quantities is calculated
by taking fully into account that light rays follow null geodesics,
as stated by general relativity. Some authors do not get this same
result because they do not take this relativistic fact into
consideration, and/or use different, or inappropriate, observational
quantities (see details in Ribeiro 2001b).} Bearing this result in mind,
it is only natural to ask whether or not a perturbed model could turn
the density decay at increasing redshift depths into a power law type
decay, as predicted, and claimed to be observed, by the scale invariant
description of galaxy clustering (Pietronero 1987; Coleman and Pietronero
1992; Pietronero et al.\ 1997; Ribeiro and Miguelote 1998; Sylos-Labini
et al.\ 1998; Pietronero and Sylos-Labini 2000).

It must be clearly understood that the effects described above appear not
simply by carrying out calculations along the null cone, but by doing
this by means of a set of observational relations whose original purpose
was power-law density profile characterization, and which turned out
to be more useful than originally envisaged. That was fully explained
in Ribeiro (1993, 1994, 1995), and it implies that if one is simply doing
calculations along the null cone it is most probable that one will find
no scaling pattern of any kind at all.\footnote{ \ See also Ribeiro
(2001b) for the pitfalls of such a simplistic approach when dealing with
the problem of the possible observational smoothness of the Universe.}
Thus, to even start considering scale invariance in relativistic cosmology
it was necessary to adapt the original analytical tools proposed by
Pietronero (1987) into a relativistic framework, and when doing this it
became clear that the chosen set of observational quantities had to have
their behaviour studied along the past null cone (Ribeiro 1992a). Therefore,
the surprising results stated in the previous paragraphs could only have
appeared through a mix of the use of a specific set of relativistically
adapted observational relations, and the realization that even at small
scales ($z < 0.1$) relativistic effects start to play an important
role in cosmology. 

The {\it apparent fractal conjecture}, as advanced by Ribeiro
(2001ab), states essentially that the observed power-law average density
profile of large-scale galaxy distribution should be a consequence of
the fact that astronomical observations are made on the backward null
cone, and, therefore, observational quantities necessary for scale
invariant characterization must be calculated accordingly. If we take
together these two considerations into account in small scales perturbed
standard models, then we should find a scaling pattern on average
density in the sense of power-law density profiles. If this hypothesis
proves, even partially, correct, many of the discrepancies between
both sides of the above mentioned debate could be immediately resolved,
and without the need of bringing into question the standard
cosmological model or even the cosmological principle. Moreover, it is
important to mention that galaxy catalogues have for some time been
showing a power-law behaviour for the density-distance correlation
(Davis et al.\ 1988; Geller 1989), and, therefore, this average density
power-law profile for the galaxy distribution could be explained by
this conjecture.

This paper seeks to prove whether or not Ribeiro's (2001ab) 
conjecture is correct, at least in a narrow sense. Our aim here
is to investigate if a perturbed standard cosmological model could show
scaling features similar to the power-law density profile predicted by
Pietronero and collaborators.

Here we show that by starting from the simplest possible cosmological
model, EdS, and carrying out a specific metric perturbation
appropriate to our needs, the conjecture is correct under an
approximation which is very reasonable if we consider the large error
margins produced by astronomical observations. We followed the
perturbative scheme proposed by Abdalla and Mohayaee (1999), where the
scale factor is expanded in power series to yield perturbative terms. 
However, in order to use this scheme we had first to derive
observational relations along the past null cone and then relate the 
results with actual observations as obtained in astronomy. Then we found
numerical solutions which show fractal like scaling features, in the
sense of Pietronero (1987).

The paper is organized as follows. In section 2 we summarize the
perturbative method used here and present the perturbed EdS
spacetime. Section 3 deals with calculating the various
observational relations in the chosen spacetime, and section 4 discusses
the numerical scheme which is used to obtain numerical solutions
for the observational quantities. Section 5 shows the numerical results
obtained, and how a scale invariant pattern appears from these results.
The paper finishes with a conclusion. 

\section{The Perturbed Metric}

Let us start with the inhomogeneous spherically symmetric metric as
proposed by Abdalla and Mohayaee (1999),
\begin{equation}
  dS^2=-dt^2+R^2(r,t) \left[ \frac{dr^2}{f^2(r)} +r^2 d\Omega^2 \right],
  \label{-1}
\end{equation}
where
\begin{equation}
  d\Omega^2=d\theta^2 + \sin^2 \theta d \phi^2, 
  \label{2}
\end{equation}
and \be f^2(r)=1-kr^2; \; \; \; \; \; k=0,\pm 1. \lb{2a} \ee
The proposal is to solve Einstein's field equations,
\begin{equation}
  R_{ab} - \frac{1}{2}g_{ab}R=-8 \pi G T_{ab},
  \label{2.1}
\end{equation}
for a perfect fluid universe with the metric above, but by means of series
expansions of the form,
\begin{equation}
    R(r,t)  =  \sum_{n=0}^{\infty} \frac{R_n(t)}{r^n}, \; \; \; \; 
    \rho(r,t)  =  \sum_{n=0}^{\infty} \frac{\rho_n(t)}{r^n}, \; \; \; \;
    p(r,t)  =  \sum_{n=0}^{\infty} \frac{p_n(t)}{r^n}. 
  \label{2.2}
\end{equation}
The zeroth order terms in this expansion are of unperturbed standard
cosmologies. Thus, $R_0(t)$ is the scale factor of the Friedmann
universe.

As a first approach to modelling a smoothed-out and averaged fractal
system in the standard cosmology, both the metric (\ref{-1}) and its
perturbation scheme, given by equations (\ref{2.2}), are well suited
for the purposes of this work, inasmuch as all previous relativistic
fractal cosmologies have so far been proposed in the LTB spacetime
(see \S 1 above). This means that equations (\ref{-1}) and (\ref{2.2})
are special cases of the LTB metric (Ribeiro 1992a), making it possible
to compare the results of this paper with the relativistic fractal
cosmologies already known. In addition, it must be noted that here we
are taking an operative definition of fractality, which refers to 
the property shown by the observed large-scale distribution of galaxies
of having an average density power-law type decay at increasing distances.
So, in this paper fractality means in fact {\it observational fractality},
in the astronomical sense, and only resembles non-analytical fractal sets
in the sense that if we define a smooth-out average density on those sets,
the properties of this average density are similar to what is found in
observational cosmology data. In other words, they are both of power-law
type ones. Therefore, under this operative definition, we can talk about
fractality, or fractal properties, in completely smooth relativistic
cosmological models, where the cosmological fluid approximation is assumed
(Ribeiro 2001b).

For flat matter dominated universe ($p=0$, $k=0$), that is, for perturbed
EdS cosmology, the metric
\be
  dS^2=-dt^2+R^2(r,t) \left( dr^2 +r^2 d\Omega^2 \right),
  \lb{1}
\ee
produces solutions of the field equations with perturbative terms that
represent growth of inhomogeneities. These, to first order, are
\begin{equation}
  R(r,t)=At^{2/3} + \left( \frac{-9C_1}{10A^2} \right)
  \frac{t^{4/3}}{r^3},
  \label{3}
\end{equation}
\begin{equation}
  \rho(r,t)= \frac{1}{6 \pi G t^2} + \left( \frac{-3C_1}{10
  \pi G A^3} \right) \frac{t^{-4/3}}{r^3},
  \label{4}
\end{equation}
where $A$ and $C_1$ are constants.
The first terms of the right hand side of equations (\ref{3})
and (\ref{4}) are of unperturbed Friedmann universe and the two 
additional terms in both equations represent the first perturbative
inhomogeneous corrections which yield growing modes. Terms corresponding
to decaying modes are ignored here. This perturbative solution reproduces
standard results, such as found in Weinberg (1972), and it corresponds to
a soft perturbation, which is almost homogeneous at large values of the
coordinate $r$ (see details in Abdalla and Mohayaee 1999). There are additional
perturbative terms which also yield growing modes, but in order to
try a first check of the possible validity of the apparent fractal
conjecture, we only need the simplest perturbative model. Therefore,
at this stage we will ignore other terms in the series.  

We shall need for later usage the time derivative of equation (\ref{3}),
\be
  \frac{\partial R}{\partial t}= \frac{2}{3}At^{-1/3}+\left(
  \frac{-6C_1}{5A^2} \right) \frac{t^{1/3}}{r^3}.
  \lb{3a}
\ee

As a final remark, while the proposed perturbation is the most
convenient for the purposes of this work, as explained above, it remains
to be seen whether or not other types of  perturbations could also be well,
or better, suited for checking the validity of the apparent fractal
conjecture. We shall not pursue this investigation here.

\section{Observational Relations Along the Past Null Cone}

The first step towards obtaining observational relations in the
spacetime given by metric (\ref{-1}) is taken by solving its 
{\it past} radial null geodesic. This astronomically important
hypersurface provides the geometrical locus for light rays that
travel towards us. It is obtained when we take $dS^2=d \theta^2=d
\phi^2=0$ in metric (\ref{-1}). In this way we obtain the following
expression, 
\begin{equation}
  \frac{dt}{dr}= - \frac{R}{f}.
  \label{5}
\end{equation}
Note that it is just a matter of convenience to write the past
radial null geodesic
above as having the radial coordinate $r$ as its parameter. In fact, both
$r$ and $t$ coordinates are functions of the null cone affine parameter
$\lambda$, which means that the equation above may also be alternatively
written as
\begin{equation}
  \frac{dt}{d\lambda}= - \frac{R}{f} \frac{dr}{d \lambda}.
  \label{6}
\end{equation}

It is rather a difficult task to obtain an analytical solution for 
the null geodesic (\ref{5}) in the flat matter dominated case ($f^2=1$),
or equivalently, to derive analytical expressions for the observational
relations along the past light cone for the perturbative solutions
(\ref{3}) and (\ref{4}). Thus, we choose an alternative approach. We
first derive all the necessary observational relations for metric
(\ref{-1}), and then solve the problem numerically to obtain solutions
corresponding to the perturbation of the metric (\ref{1}), as
given by equations (\ref{3}) and (\ref{4}). Such a procedure will
eventually allow us to obtain the desired observational relations,
although in numerical form.  To pursue this path, we should start by
deriving the redshift in the geometry given by equation (\ref{-1}).

The general expression for the {\it redshift}, in {\it any} spacetime,
is given by (see, {\it e.g.}, Ellis 1971)
\begin{equation}
  1+z=\frac{ { \left( u^a k_a \right) }_{ \scriptstyle{\rm source}}}{{
  \left( u^a k_a \right) }_{\scriptstyle{\rm observer}}},
  \label{6.1}
\end{equation}
where $u^a$ is the 4-velocity of source and observer and $k^a$ is the
tangent vector of the null geodesic joining them. If source and observer
are comoving, then $u^a= \delta^a_0$, and equation (\ref{6.1}) becomes
\begin{equation}
  1+z=\frac{ { \left( dt/d \lambda\right) }_{ \scriptstyle{\rm
  source}}}{{ \left( dt/d \lambda\right) }_{\scriptstyle{\rm
  observer}}},
  \label{10}
\end{equation}
since, $g_{00}=-1$, and, by definition, $k^0=dt/d \lambda$.

Finding $(dt/d \lambda )$ at both source and observer requires the use
of an indirect method, which will be described as follows. We start with
the Lagrangian for the radial metric,
\begin{equation}
  {\cal L}= - {\left( \frac{dt}{d \lambda} \right) }^2 +
  \frac{R^2(r,t)}{f^2(r)} {\left( \frac{dr}{d \lambda} \right) }^2.
  \label{12}
\end{equation}
The Euler-Lagrange equations of motion
\be
  \frac{d}{d \lambda} \frac{ \partial {\cal L}}{ \partial \dot{q}} -
  \frac{ \partial {\cal L}}{ \partial q} = 0,
  \lb{13}
\ee
can be applied to equation (\ref{12}), yielding
\be
  \frac{d \dot{t}}{d \lambda} + \left( \frac{R}{f^2} \frac{\partial
  R}{\partial t} \right) {\dot{r}}^2 =0,
  \lb{16}
\ee
\be
  \frac{d \dot{r}}{d \lambda} + \frac{1}{R} \left( \frac{\partial
  R}{\partial r} - \frac{R}{f} \frac{df}{dr} \right) {\dot{r}}^2 +
  \left( \frac{2}{R} \frac{\partial R}{\partial t} \right) \dot{r}
  \dot{t} =0,
  \lb{17}
\ee
where the dot means derivative with respect to the affine parameter
$\lambda$. If we use the null geodesic (\ref{6}) in equations 
(\ref{16}) and (\ref{17}), they can both be integrated once. The results
may be respectively written as
\be
 \dot{t}= { \left[ \int \left( \frac{1}{R} \frac{ \partial R}{ \partial t}
 \right) d \lambda + b_1 \right] }^{-1},
 \lb{19}
\ee
\be
 \dot{r}= { \left[ \int \left( \frac{1}{R} \frac{ \partial R}{ \partial
 r} - \frac{1}{f} \frac{df}{dr}- \frac{2}{f} \frac{ \partial R}{
 \partial t} \right) d \lambda + b_2\right] }^{-1},
 \lb{20}
\ee
where $b_1$ and $b_2$ are integrations constants.

To find those constants, let us now write a 2-surface displacement, with
$t$ and $\phi$ constants, of metric (\ref{-1}),
\be
  dS^2= \frac{R^2}{f^2} \left( dr^2 + f^2 r^2 d \theta^2 \right).
  \lb{20.1}
\ee
We shall now require the metric to be regular at the spatial origin, that is,
as $r \rightarrow 0$ the metric must be Euclidean. Therefore, $f^2
\rightarrow 1$, $R^2/f^2 \rightarrow \mbox{constant}$, as $r \rightarrow 0$.
In other words, we are requiring that metric (\ref{-1}) should obey the
{\it central regularity condition} (Bonnor 1974; Ribeiro 1993;
Humphreys, Matravers and Marteens 1998),
\be
  \lim_{r \rightarrow 0} R =1.
  \lb{21}
\ee

Now, if we re-substitute solutions (\ref{19}) and (\ref{20}) back into the
null geodesic (\ref{6}) we obtain,
\be
  \left[ \int \left( \frac{1}{R} \frac{ \partial R}{ \partial
  r} - \frac{1}{f} \frac{df}{dr} - \frac{2}{f} \frac{ \partial R}{
  \partial t} \right) d
  \lambda + b_2\right] = - \frac{R}{f} \left[ \int \left( \frac{1}{R}
  \frac{ \partial R}{ \partial t}
  \right) d \lambda + b_1 \right].
  \lb{21.1}
\ee
This equation is valid for any $\lambda$, including at the origin, where
the observer is located. From now on we will be labelling the event of
observation as $r=\lambda=0$. So, considering the regularity condition
(\ref{21}), equation (\ref{21.1}) may be written as,
\be
  b_2=- b_1
  \lb{22}
\ee
If we now consider the same regularity conditions, then equation
(\ref{19}) becomes, 
\be
  { \left[ \frac{dt}{d \lambda} \right] }_{\lambda=0} = \frac{1}{b_1}.
  \lb{22.1}
\ee
Inasmuch as, we are interested in incoming light rays, {\it i.e.}, our
model deals with photons along the {\it past} light cone, it is natural
to choose $b_1=-1$ as the value for this constant. Therefore,
\be
  b_1=-1, \: \; \; \; \; \Rightarrow \: \; \; \; \; b_2=1,
  \lb{23}
\ee
and we may write equations (\ref{19}) and (\ref{22.1}) as, 
\be
  \frac{dt}{d \lambda}= { \left[ \int \left( \frac{1}{R}
  \frac{ \partial R}{ \partial t} \right) d \lambda -1 \right]
  }^{-1},
  \lb{24}
\ee
\be { \left[ \frac{dt}{d \lambda} \right] }_{\lambda=0} = -1.
\lb{24.1} \ee
If we now define an auxiliary term, named as {\it I-term}, as being
given by,
\be
 I \equiv \int \left( \frac{1}{R}\frac{ \partial R}{
 \partial t} \right) d \lambda,
 \lb{25}
\ee
equations (\ref{24}) and (\ref{24.1}) allow us to re-write the redshift
(\ref{10}) as follows,
\be
  z=\frac{I}{1-I}.
  \lb{26}
\ee

We still have to calculate the I-term in order to evaluate the redshift,
and this can be done as follows. Considering equations (\ref{23}) and
(\ref{25}) we may re-write equation (\ref{21.1}) as,
\be
  \left\{ \int \left[ \frac{1}{R} \frac{ \partial R}{ \partial
    r} - \frac{1}{f} \frac{df}{dr} - \frac{2}{f} \frac{ \partial
    R}{ \partial t} \right] d \lambda + 1 \right\} = - \frac{R}{f}
    \: \left( I - 1 \right).
  \lb{a2}
\ee
Thus, considering equations (\ref{25}) and (\ref{a2}), equations (\ref{19})
and (\ref{20}) may be re-written as,
\be
  \frac{dt}{d \lambda} = \frac{1}{I-1},
  \lb{a3}
\ee
\be
  \frac{dr}{d \lambda} = \frac{f}{ \left( 1-I \right) R}.
  \lb{a4}
\ee
{From} equation (\ref{25}) it is easy to see that,
\be 
  \frac{d I}{dr} = \frac{d}{dr} \int \frac{1}{R} \frac{ \partial R}{
  \partial t} \frac{d \lambda}{dr} \; dr,
  \lb{a4.1}
\ee
and, by substituting equation (\ref{a4}) in the expression above, we finally
obtain,
\be
  \frac{d I}{dr} = \left( \frac{ 1-I }{f} \right) \frac{
  \partial R}{\partial t}.
  \lb{a5}
\ee

The solution of the first order ordinary differential equation
(\ref{a5}) allows us to calculate the I-term and, as a consequence,
the redshift, as given by equation (\ref{26}). When $r \rightarrow 0$,
$f=1$, $R=1$ and $I=0$, $(dI/dr)=0$. Remembering the perturbed solution
(\ref{3}), we are facing again a differential equation whose analytical
solution is difficult, if not impossible, to find. 

The other observational relations relevant to the problem under
consideration can be straightforwardly calculated now. The {\it observer
area distance}, or simply {\it area distance},\footnote{ \ This definition
of distance is the same as Weinberg's (1972) {\it angular diameter
distance}, and Kristian and Sachs' (1966) {\it corrected luminosity
distance}. A detailed discussion about distances in cosmology as applied
to the problem discussed in here can be found in Ribeiro (2001b).} as
defined by Ellis (1971) for {\it any} spacetime, is given by
\be ({d_A})^2 = \frac{dA_0}{d \Omega_0}, \lb{34.1} \ee
where $d \Omega_0$ is the solid angle element for constant $r$, and
$dA_0$ is the cross sectional area for this solid angle (see Ellis 1971).
For metric (\ref{1}) we have,
\be d \Omega_0 = d \theta \sin \theta d \phi, \; \; \; \; \; \;
    dA_0 = R^2 r^2 \sin \theta d\theta d \phi.
    \lb{34.2}
\ee
Therefore, the area distance is given by the following expression,
\be {d_A}= r R. \lb{35} \ee

The {\it luminosity distance} is obtained from the area distance by
means of Etherington's {\it reciprocity theorem} (1933; see also
Ellis 1971; Schneider, Ehlers and Falco 1992), which relates both
distance definitions through the expression
\be { ( d_\ell )}^2 = ({d_A})^2 {( 1+z )}^4. \lb{35.1} \ee
So, for the spacetime (\ref{1}) we have
\be
  d_\ell = r R \; { \left( 1 +z \right) }^2 =
	   \frac{r R}{ { \left( 1-I \right) }^2 }.
  \lb{36}
\ee

The general expression for {\it number counting} in any cosmological
model at a point $P$ down the null cone is given by (Ellis 1971),
\be dN= ({d_A})^2 d \Omega_0 { \left[ n \left( -k^a u_a \right) \right]
    }_P \; d \lambda. 
    \lb{36.1}
\ee
Here $n$ is the number density of radiating sources per unit proper
volume. Considering equation (\ref{6}) and that $k^0=dt/d \lambda$ for
comoving source, equation (\ref{36.1}) becomes,
\be dN= 4 \pi n r^2 R^3 dr, \lb{37} \ee
where we have performed an integration over all solid angles. If we now
make the assumption that all sources are galaxies, with approximately
the same average mass, then
\be n=\frac{\rho}{M_g}, \lb{37.1} \ee
with $M_g \approx 10^{11}M_{\odot}$ being the rest mass of an average
galaxy, and $\rho$ comes from equation (\ref{4}), we obtain another
differential equation to be solved numerically, 
\be 
  \frac{dN}{dr}=\frac{4 \pi}{M_g} \rho \; r^2 R^3. \lb{38}
\ee

Finally, to discuss fractality, in the sense of a smoothed-out and
averaged scale invariant system with a decaying power-law profile for
the average density, we also need the {\it observed volume},
here defined as
\be V=\frac{4}{3} \pi { ( d_\ell )}^3, \lb{58} \ee
and the {\it observed average density},
\be \langle \rho \rangle = \frac{M_g N}{V}. \lb{59} \ee
Ribeiro (1995) showed that in an unperturbed EdS model the following
relations holds, 
\be \Gamma^\ast=\frac{\langle \rho \rangle}{M_g}=\langle n \rangle.
    \lb{new1}
\ee
Therefore, Pietronero's (1987) conditional average density
$\Gamma^\ast$ is equal to the observed average number density in an
unperturbed EdS model. Since, these two quantities are fundamental in
the characterization of a single fractal structure (Ribeiro and
Miguelote 1998),
they can also be expected to be equally fundamental in the fractal
characterization of the perturbed model considered here. 

\section{Numerical Problem}

We have seen in the previous section that in order to obtain observational
relations for flat matter dominated universe, we need to solve three
differential equations: the past radial null geodesic (\ref{5}), the
I-term (\ref{a5}) and the cumulative source number count (\ref{38}).
However, in obtaining the I-term and source count we need first to solve
the null geodesic, as both depend on the scale factor $R(r,t)$. In other
words, integrating the null cone produces a solution given by the
function $t=t(r)$, which is necessary for integrating the I-term and
source count along the null cone.

The integration procedure outlined above can be algorithmically
expressed as follows. We  start with the initial conditions $r_1$, $t_1$,
$I_1$, $N_1$, use the first two to find $R_1$, $\rho_1$, $[\partial R/ 
\partial t]_1$, by means of expressions (\ref{3}) and (\ref{4}), use
some numerical code for solving ordinary differential equations to advance
one step and finally find $t_2$, then $I_2$, $N_2$. As $r_2$ is known in
advance, since it is the independent variable, the newly found values $t_2$,
$I_2$,$N_2$ are used to repeat the cycle until we finish the
integration, in $r_n$. Although that amounts to a simple numerical procedure,
some care is needed in order to make sure we will be using the values
obtained in the integration of the null geodesic to feed the evaluation
of functions (\ref{3}) and (\ref{4}). In other words, if $t_i$ is an
intermediary value $(i=1,\ldots,n)$, obtained numerically, of the null
geodesic $t=t(r)$, then $t_i$ must be used to find $R_i$, $\rho_i$,
$[\partial R/ \partial t]_i$, which are then used to find $t_{i+1}$,
$I_{i+1}$, and $N_{i+1}$, and so on.

The initial values pose a problem: if we start the integration at $r=0$,
according to equations (\ref{2.2}) we will face a singularity at the
origin. To avoid this difficulty we will assume a flat and Euclidean
spacetime from $r=0$ up to $r=\varepsilon$, where $\varepsilon$
will be as small as necessary. As seen above, previous studies have
shown that observational departures from spatial homogeneity occur at $z
\approx 0.04$, which means $\sim$ 160 Mpc. Therefore, it is reasonable
to assume a Euclidean spacetime up to $\sim$ 100 Mpc, or $z \approx 0.03$,
which means taking $\varepsilon = 0.1 \lb{a7} $
as the initial integration value. Here we will be taking distances in
Gpc and units such that $c=G=1$. In these units $M_g=10^{11}M_{\odot} =
4.787591 \times 10^{-12}$ and $H_0=0.250173$ for the value of the Hubble
constant of 75 km s$^{-1}$ Mpc$^{-1}$ in the usual units.

The assumption above is completely coherent with the central regularity
condition (\ref{21}), but in fact introduces the notion that there is a
Euclidean to non-Euclidean interface at $r=\varepsilon$. This means that
we need to find initial values at $r=\varepsilon$ to start the
integration. Then, at this interface, the null geodesic reduces to
\be t=-r+t_0, \lb{60} \ee
which implies the following initial values,
\be \left\{ \begin{array}{lll}
	    R & = & 1, \\
	    t & = & -\varepsilon + t_0, \\
	    r & = & \varepsilon,
	    \end{array}
    \right.
    \lb{60.1}
\ee
where $t_0$ is the label given by the time coordinate at the present epoch.
Therefore, equation (\ref{60}) implies that the event of observation, that
is, the ``here and now'', is labeled by $r=\lambda=0$, and $t=t_0$.
Since the time elapsed since the big bang singularity hypersurface is
the same for all observers in the standard cosmologies, we may take
$t_0$ to be the same as in the unperturbed EdS model, that
is, 
\be t_0=\frac{2}{3H_0}. \lb{time} \ee

The initial values (\ref{60.1}), once applied to equation (\ref{3}),
allows us to find an expression linking the two constants $C_1$ and
$A$,
\be 
  C_1= \frac{10}{9} A^2 \varepsilon^3 \left[ A {\left( \frac{2}{3H_0} -
  \varepsilon \right) }^{2/3} -1 \right] {\left( \frac{2}{3H_0} -
  \varepsilon \right) }^{-4/3}.
  \lb{61}
\ee

In the flat and Euclidean region close to the origin, the universal
density is assumed to be constant, whose value should be the critical
density for Friedmann universe, that is, the value of the local density
for a EdS universe at present time. By using this
requirement in equation (\ref{4}), we get,
\be
  \rho_0=\frac{3{H_0}^2}{8 \pi}= \frac{1}{6 \pi} {\left( \frac{2}{3H_0}
  -\varepsilon \right) }^{-2} - \frac{3C_1}{10 \pi A^3 \varepsilon^3}
  {\left( \frac{2}{3H_0} -\varepsilon \right) }^{-4/3},
  \lb{61.1}
\ee
or, putting $C_1$ in evidence,
\be 
 C_1= \frac{15 A^3 {H_0}^3 \varepsilon^4}{4 \left( 2-3 \varepsilon H_0
      \right)} {\left( \frac{2}{3H_0} -\varepsilon \right) }^{4/3}.
 \lb{62}
\ee
Equations (\ref{61}) and (\ref{62}) provide conditions for the two
unknown constants $C_1$ and $A$ to be expressed in terms of the Hubble
constant. We thus obtain,
\be
 A= \frac{8}{ \left[8-3 \varepsilon H_0 \left(2-3 \varepsilon H_0
    \right) \right] } {\left( \frac{2}{3H_0} -\varepsilon \right)
    }^{-2/3},
 \lb{63}
\ee
\be
 C_1 = \frac{1920 {H_0}^3 \varepsilon^4}{ \left(2-3 \varepsilon H_0
       \right) {\left[8-3 \varepsilon H_0 \left(2-3 \varepsilon
       H_0\right) \right] }^3} {\left( \frac{2}{3H_0} -\varepsilon
       \right) }^{-2/3}.
 \lb{64}
\ee
A power series expansion for $\varepsilon$ in equation (\ref{63}) yields
\be
 A = { \left( \frac{3H_0}{2} \right) }^{2/3} + O(\varepsilon),
 \lb{64a}
\ee
while in equation (\ref{64}), a similar expansion produces,
\be
 C_1 = \frac{15}{8} { \left( \frac{3H_0}{2} \right) }^{2/3} {H_0}^3
       \varepsilon^4 + O(\varepsilon^5).
  \lb{64b}
\ee
Thus, if $\varepsilon$ is too small, meaning a too small flat
region, then $A$ remains a nonzero constant, and $C_1$ becomes negligibly
small. In such a case the perturbative terms in equations (\ref{3}) and
(\ref{4}) will vanish.

Finally, for number count (\ref{38}), up to $r=\varepsilon$ we have
\be \frac{dN}{dr} = \frac{4 \pi}{M_g} \rho_0 r^2, \; \; \; \; \; 
    \Longrightarrow \; \; \; \; \; N(\varepsilon)=\frac{{H_0}^2
    \varepsilon^3}{2 M_g}.
    \lb{64.1}
\ee
This last equation implies that,
\be { \left. \langle \rho \rangle \right| }_{r=\varepsilon}
    = \rho_0,
    \lb{64.2}
\ee
as it should.

\section{Numerical Solutions}

The previously defined procedure for computing observational relations
in a perturbed model was carried out by means of a simple numerical code
written in {\sc fortran 77}. The results are better or worse depending
on the size of the flat region, {\it i.e.}, depending on the value of
$\varepsilon$. For very small values, due to the fourth power for
$\varepsilon$, as seen in equation (\ref{64b}), the
perturbation tends to vanish and the results are similar to those found
years ago by Ribeiro (1992b, 1993, 1994, 1995). For larger values of
$\varepsilon$ the form of the curve showing the average density $\langle
\rho \rangle$ as plotted against the luminosity distance $d_\ell$ does
approach a linear behaviour in a log log scale, meaning a power law
decay for the observed average density along the past null cone. Such
a behaviour means that our perturbed EdS cosmological model can be
approximately described by a decaying power-law like distribution, a
result which validates, in an approximate manner, the apparent fractal
conjecture, at least by means of the perturbation scheme used here.

Figure \ref{fig1} shows the best numerical result obtained, and there
\begin{figure}[p]
\begin{center}
\input{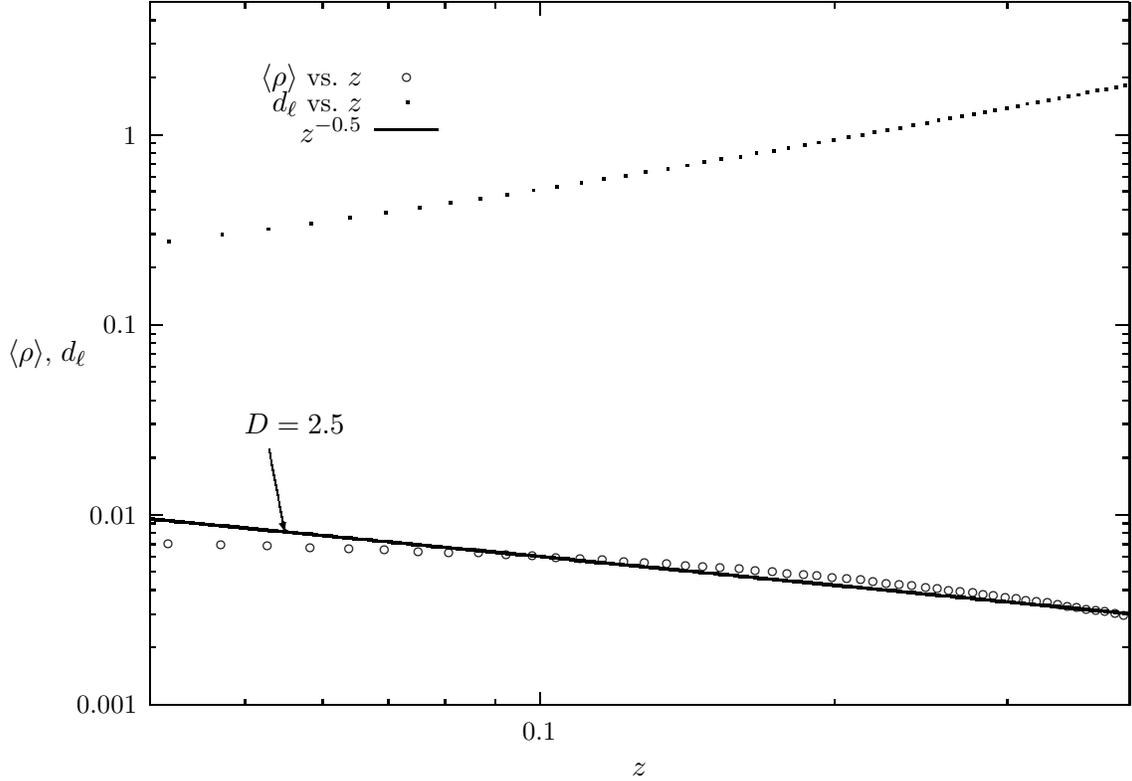}
\caption{This figure shows a plot in the redshift range $0.04 \le z \le
0.4$, containing numerical results of the observational quantities of
interest, namely, the observed average
density $\langle \rho \rangle$, the luminosity distance $d_\ell$, and
the redshift $z$. These observables were calculated along the past
null cone, and by using numerical solutions of the differential
equations (\protect\ref{5}), (\protect\ref{a5}), and (\protect\ref{38}).
The value used to limit the inner flat region was $\varepsilon=0.1$,
meaning that its range is of 100 Mpc, or $z \approx 0.03$. One can clearly
see above that the decay of $\langle \rho \rangle$ against an increasing
$z$ is approximately linear (open circles), as expected for a scale
invariant system. The straight line fitted to the points has power $-0.5$,
which means that the decay of the average density has fractal dimension
$D=2.5$. So, despite the simplicity of the model the fractal dimension
has a value not too different from what is observed. The figure also
presents a plot for $d_\ell$ vs.\ $z$ (dots), showing that the linearity
of the Hubble law, {\it i.e.}, the distance-redshift relation, is well
retained within the integration range. This proves that the apparent
fractal conjecture allows for a fractal like power-law decay of the
average density to co-exist with the Hubble law, this being no
``paradox''.}
\lb{fig1}
\end{center}
\end{figure}
one can see that the property of a smoothed-out and averaged fractal system
of exhibiting a linear decay of its average density, due to its power law
feature, appears approximately. If we consider that astronomical data usually
have large error margins, and that spatially flat LTB models are known at
providing not so good fractal modelling (Ribeiro 1993), the results shown
in figure \ref{fig1} can be considered as quite good. Therefore, in spite of the
simplicity of the model we have succeeded in finding an observational
scaling behaviour in the EdS cosmology, and with a fractal dimension not
too different from the expected value foreseen by observations (see, {\it
e.g.}, Ribeiro and Miguelote 1998, Sylos-Labini et al.\ 1998). 

Previous studies with this kind of fractal modelling in LTB spacetimes
showed that spatially flat models do not provide very good results,
while open models provide the best ones (Ribeiro 1993, 1994; see also
Humphreys, Matravers and Marteens 1998). Therefore, in view of this it
is reasonable to conclude that the best results regarding the validity
of the apparent fractal conjecture should come from perturbed open
Friedmann cosmologies.

The situation can certainly be improved by either modifying the type of
perturbation, and also (possibly more important) if we better accommodate
the observed nearby matter distribution, seen here as just flat geometry.

\section{Conclusion}

In this paper, we have used a perturbative model of the Einstein-de
Sitter cosmology to test the validity of the hypothesis known
as ``the apparent fractal conjecture'' (Ribeiro 2001ab), which states that
the observed power-law average density profile decay, derived from
observations of large-scale distribution of galaxies, appears
when observational quantities relevant for scale invariant
characterization are calculated along the past light cone in universe
models with small scales perturbations. These are the quantities which
should be directly compared to astronomical observations. We have
started with the perturbative method proposed by Abdalla and Mohayaee
(1999), where the scale factor is expanded in power series to yield
perturbative terms, and then derived observational relations necessary
for checking the validity of this conjecture in a perturbed spacetime,
which is almost homogeneous at large values of the radial coordinate.
The observational quantities derived, namely the redshift, area distance,
luminosity distance, number counting, observed volume and average density,
are all dependent on the solution of three ordinary first-order
differential equations, which cannot be integrated analytically.
Consequently, we have produced a numerical scheme for integrating these
equations, namely the past radial null geodesic, the I-term, necessary for
redshift evaluations, and the integrated source number count. We have
found numerical solutions which show that the observed dust distribution
in this perturbed Einstein-de Sitter cosmological model can be approximately
described by a decaying power-law like density profile, that is, by a scale
invariant, smoothed-out and averaged galaxy distribution, which is 
characterized by an unique and non-integer dimension. In other words, by
a single fractal system. This result proves that the apparent
fractal conjecture is correct, at least under the specific perturbative
approach adopted here.

As consequences of our results, it is important to point out that it
remains to be seen whether or not the conjecture is also valid in
different, or more general, small scales perturbations to standard
cosmologies. However, even if we only consider our simple perturbative
approach to the problem, as described above, it is clear that the
``fractal debate'' currently underway (see \S 1) does not necessarily
need to continue developing in antagonistic viewpoints. Our results
suggest that the observed scale invariant pattern may be understood from within
the theoretical context of standard Friedmann cosmology,\footnote{ \ A
similar conclusion has recently been reached by Joyce et al.\ (2000).}
and the apparent fractal conjecture may provide a starting point for
developing the conceptual framework aimed at including scaling ideas,
and scaling related data, into main stream cosmological research. In
such a case the cosmological principle and the possibility of an
infinity scale invariant, or fractal, system could, perhaps, be
reconciled into a single theoretical framework. In addition, if
the observed power-law density profile does appear with perturbative
solutions of Friedmann cosmologies, then we may speculate that the
observed scale invariant distribution of galaxies may be considered
as direct observational evidence of primordial density fluctuations
in the Universe, perhaps in a similar way as anisotropies of the
cosmic microwave background radiation.

\begin{flushleft}
{\Large \bf Acknowledgements}
\end{flushleft}
We would like to thank L.\ Pietronero for reading the original
manuscript and for helpful comments and remarks. We are also grateful to
L.\ Amendola for discussions and for pointing out a mistyping. EA wishes
to thank CNPq for partial support, and FAPESP for the thematic project
97/06499-2. MBR is grateful to FUJB-UFRJ for partial support.
\begin{flushleft}
{\Large \bf References}
\end{flushleft}
\begin{description}
\item Abdalla, E., Afshordi, N., Khodjasteh, K., and Mohayaee,
      R.\ 1999, Astron.\ Astrophys., 345, 22
\item Abdalla, E., and Mohayaee, R.\ 1999, Phys.\ Rev.\ D, 59,
      084014, astro-ph/9810146
\item Bonnor, W.\ B.\ 1974, Monthly Not. Royal Astron.\ Soc., 167, 55
\item Coleman, P.\ H., and Pietronero, L.\ 1992, Phys.\ Rep., 213, 311
\item Coles, P.\ 1998, Nature, 391, 120
\item Davis, M.\ et al.\ 1988, Astrophys.\ J., 333, L9 
\item de Vaucouleurs, G.\ 1970, Science, 167, 1203
\item Ellis, G.\ F.\ R.\ 1971, General Relativity and Cosmology,  Proc.\
      of the International School of Physics ``Enrico Fermi'', R.\ K.\ Sachs,
      New York: Academic Press, 1971, 104
\item Etherington, I.\ M.\ H.\ 1933, Phil.\ Mag., 15, 761; reprinted in
      Gen.\ Rel.\ Grav., in press (2001)
\item Geller, M.\ 1989, 
      Astronomy, Cosmology and Fundamental Physics, M. Caffo et al.,
      Dordrecht: Kluwer, 1989, 83 
\item Humphreys, N.\ P., Matravers, D.\ R., and Marteens, R.\ 1998,
      Class.\ Quantum Grav., 15, 3041, gr-qc/9804025
\item Joyce, M., Anderson, P.\ W., Montuori, M., Pietronero, L., and
      Sylos-Labini, F.\ 2000, Europhys.\ Lett.\  50, 416, astro-ph/0002504
\item Kihara, T., Saki, K.\ 1970, Publ.\ Astron.\ Soc.\ Japan, 22, 1
\item Kristian, J., and Sachs, R.\ K.\ 1966, Astrophys.\ J., 143, 379
\item Longair, M.\ S.\ 1995, The Deep Universe, Saas-Fee Advanced Course
      23, B.\ Binggeli and R.\ Buser, Berlin: Springer, 1995, 317
\item Mart\'{\i}nez, V.\ J.\ 1999, Science, 284, 445
\item Mandelbrot, B.\ B.\ 1983, The Fractal Geometry of Nature,
      New York: Freeman
\item Matravers, D.\ R.\ 1998, Proc. of the Intl.\ Seminar on Mathematical
      Cosmology, M.\ Rainer and H-J.\ Schmidt, Singapore: World Scientific,
      gr-qc/9808015
\item Peebles, P.\ J.\ E.\ 1980, The Large-Scale Structure of the Universe,
      Princeton University Press
\item Pietronero, L.\ 1987, Physica A, 144, 257
\item Pietronero, L., Montuori, M., and Sylos-Labini, F.\ 1997, Critical
      Dialogues in Cosmology, N.\ Turok, Singapore: World Scientific, 1997, 24 
\item Pietronero, L., and Sylos-Labini, F. 2000, astro-ph/0002124
\item Ribeiro, M.\ B.\ 1992a, Astrophys.\ J., 388, 1 
\item Ribeiro, M.\ B.\ 1992b, Astrophys.\ J., 395, 29 
\item Ribeiro, M.\ B.\ 1993, Astrophys.\ J., 415, 469 
\item Ribeiro, M.\ B.\ 1994, Deterministic Chaos in General Relativity,
      D.\ Hobbil, A.\ Burd, and A.\ Coley, New York: Plenum Press, 1994, 269
\item Ribeiro, M.\ B.\ 1995, Astrophys.\ J., 441, 477, astro-ph/9910145 
\item Ribeiro, M.\ B.\ 2001a, Fractals, in press, gr-qc/9909093 
\item Ribeiro, M.\ B.\ 2001b, Gen.\ Rel.\ Grav., in press, astro-ph/0104181 
\item Ribeiro, M.\ B., and Miguelote, A.\ Y.\ 1998, Brazilian J.\ Phys., 28,
      132, astro-ph/9803218
\item Schneider, P., Ehlers, J., and Falco, E.\ E.\ 1992, Gravitational
      Lenses, Berlin: Springer
\item Sylos-Labini, F., Montuori, M., and Pietronero, L.\ 1998, Phys.\ Rep.,
      293, 61, astro-ph/9711073
\item Turok, N.\ (editor) 1997, Critical Dialogues in Cosmology, Singapore:
      World Scientific, 1997
\item Weinberg, S.\ 1972, Gravitation and Cosmology, New York: Wiley
\item Wertz, J.\ R.\ 1970, Newtonian Hierarchical Cosmology, PhD thesis
     (University of Texas at Austin, 1970)
\item Wertz, J.\ R.\ 1971, Astrophys.\ J., 164, 227
\item Wu, K.\ K.\ S., Lahav, O., and Rees, M.\ J.\ 1999, Nature, 397, 225,
      astro-ph/9804062
\end{description}
\end{document}